\documentclass[prb,twocolumn,showpacs,amsmath,floatfix]{revtex4}
\usepackage{graphicx}
\usepackage{dcolumn}
\usepackage{bm}
\begin{document}
%%%%%%%%  list of  our own commands     %%%%%%
\newcommand{\be}{\begin{equation}}
\newcommand{\ee}{\end{equation}}
\newcommand{\ba}{\begin{eqnarray}}
\newcommand{\ea}{\end{eqnarray}}
\newcommand{\vk}{{\bf k}}
\newcommand{\vq}{{\bf q}}
\newcommand{\vp}{{\bf p}}
\newcommand{\vx}{{\bf x}}
\newcommand{\half}{\frac{1}{2}}
%%%%%%%%%%%%%%%%%%%%%%%%%%%%%%%%
\title{ A path integral approach to Anderson-Holstein model}
\author{Hyun C. Lee}
\email{hyunlee@sogang.ac.kr}
\affiliation{Department of Physics and Basic Science Research Institute, Sogang University,
 Seoul, 121-742, Korea}
\author{Han-Yong Choi}
\affiliation{Department of Physics,
BK21 Physics Research Division, Institute of Basic Science, and CNNC, \\
Sung Kyun Kwan University, Suwon, 440-746, Korea.}
\date{\today}
%%%%%%%%%%%%%%%%%%%%%%%%%%%%%%%%%%%%%%%%%%%%%%%%%%%%%%%%
\begin{abstract}
The Anderson-Holstein  model is studied in the framework of the semiclassical approximation.
Analytic results for Kondo temperature renormalized by weak electron-phonon interaction and 
for phonon Green function are obtained, and they are interpreted from the viewpoint of dynamical mean field theory.
Especially the isotope effect of the  effective electron mass is discussed in the presence of strong electron correlation.
The results are also compared with those by numerical renormalization group and other related works, and they are consistent
with each other in their common domain of validity.
\end{abstract}
%%%%%%%%%%%%%%%%%%%%%%%%%%%%%%%%%%%%%%%%%%%%%%%%%%%%%%%%%%%
\maketitle
%%%%%%%%%%%%%%%%%%%%%%%%%%%%%%%%%%%%%%%%%%%%%%%%%%%%%%%%%%%%
\section{Introduction}
Both electron correlation and lattice dynamics are important in understanding the
physical properties of many condensed matter systems.\cite{orbital}
But the {\it interplay} between them has not 
been studied in detail owing to the lack of reliable calculation methods.
Conceptually this problem is difficult to handle due to the absence of
suitable expansion parameters.

Impurity problems have been relatively well studied compared to   problems on lattice, mainly
thanks to diverse analytic and numerical nonperturbative techniques.
With the recent advent of the dynamical mean field theory (DMFT) \cite{dmft}
various lattice problems can be mapped (or approximated) to  impurity problems.
The mapping can be justified rigorously in the limit of infinite spatial dimensions. 
Once the mapping to a certain impurity problem is accomplished 
the powerful techniques of impurity problems can be employed to understand the lattice problems.

In the perspective of DMFT the problem of the interplay between electron correlation and lattice dynamics can be mapped
to an impurity problem with both electron and phonon degrees of freedom.
One of the simplest model with both degrees of freedom is the Anderson-Holstein (AH) model.\cite{hewson,jpc}
AH model is a single-impurity Anderson model with a linear coupling to a local phonon mode as in the
Holstein model.\cite{holstein}

Our goal is to understand the physical properties of the AH model in the light of the interplay of electron correlation
and lattice dynamics.
Recently the numerical renormalization group (NRG) method has been successfully applied 
to  AH model\cite{hewson,jpc}, and almost exact results on the 
electron and phonon spectral functions have been obtained. 
However, it is still desirable  to develop an analytic scheme in spite of its approximate nature
since it helps us  to understand the underlying physics in more clear and intuitive way.
Of course, a systematic analytic approach to a model like AH model
does not exist in general.
But by restricting our focus to a certain parameter regime we can develop
an approximation scheme which captures the essential features of the interplay between electron
correlation and phonons.
We will be mostly interested in the regime where the electron correlation effects are much
\textit{stronger} [large local Coulomb repulsion $U$] than the electron-phonon coupling.
If the electron-phonon coupling is omitted (Anderson model) it is well known that the ground 
state is the Kondo singlet state.
Our main focus is on  understanding  how the spin fluctuation processes responsible 
for the  Kondo ground state are influenced by  the {\it weak} electron-phonon interaction.
We have developed a {\it semiclassical approximation} scheme in the context of path integral approach where 
 the spin fluctuations are represented  by  instantons ( or kinks). \cite{hamann}
In this scheme  phonons can be naturally incorporated and the features of interplay
can be clearly exhibited.

Isotope effect is   a suitable  physical property  for addressing the issue of the interplay of electron correlation and phonons.
In conventional metals, where the Midal-Eliashberg (ME) theory\cite{me} 
of electron-phonon applies, the isotope effect appears to be very small.
This is because ME theory  keeps only the diagrams of the zeroth order of $\Omega/t$, 
where $\Omega$ is  phonon frequency and $t$  is the bandwidth of conduction electron.\cite{deppeler2}
The omission of diagrams of higher order in $\Omega/t$ is justified by  the Migdal theorem on the electron-phonon vertex function,
which is the key element of ME theory.
Presumably the ME theory cannot be employed for the systems with strong electron correlations 
since  vertex corrections are not expected to be small  for such systems in general.
We can  compute the isotope effect for the strongly correlated systems by studying how the Kondo temperature of impurity model
is modified by (local) electron-phonon interaction [see Eq.   (\ref{finalisotope})].

There is a caveat in directly translating the  results obtained for impurity model to those of the 
corresponding lattice model.
The lattice model with   Lorentzian density of state (DoS) is formally identical with  the impurity model since there 
is no need for solving self-consistency relation.
Thus for the lattice model with the  Lorentzian DoS the results obtained for the impurity model can be directly applied.
However, the Lorentzian DoS is not physical in many aspects, but it is expected to be valid for 
the systems with finite DoS at Fermi surface.\cite{dmft}
The constraint of finite DoS at Fermi surface restricts the applicability of our results to the metallic regime.
For the lattice Hubbard model this implies that the system should be  away from half-filling or 
$U_{c2} > U > \pi \Delta$ at half-filling [see Ref. [\onlinecite{dmft}] for details]. In addition,
our results are not relevant for the insulating regime such as small polaron regime.\cite{polaron}

We  point out that Deppeler and Millis addressed the similar  issue 
using the adiabatic expansion of  DMFT without assuming any specific ground state.\cite{deppeler2,deppeler1}
They  identified a new class of electron-phonon
diagrams missed in previous studies but they treated  electron correlations
in the scheme of Hatree-Fock approximation. 
Our approach treats the strong electron correlation more accurately, thus it provides the more 
accuruate electron correlation dependences of  physical properties.

In a single band model without extra orbital degrees of freedom, the Holstein phonon interaction acts in the 
charge channel, while the electron correlation effect, represented by local Coulomb repulsion, acts via the spin channel
at low energy. Due to this apparent ``spin-charge separation", the interplay of electron correlation and electron-phonon interaction 
first appears in the second order expansion with respect to the phonon field in our treatment.
For a multiband model where the orbital degrees of freedom is active electron-phonon dynamics becomes much richer
as exemplified by the phase diagrams of colossal magnetoresistance (CMR) materials.\cite{orbital,urushibara}
We have compared our results with those of the spin-polarized Jahn-Teller model studied in the context of 
CMR manganites.\cite{nagaosa} The only formal difference between two models lies in the structure of the coupling of phonon modes,
but this difference brings about  drastic differences  in physical properties such as isotope effects.

Our main results are the following:
(1) the electron-phonon interaction {\it increases} the Kondo energy scale in the regime of strong electron correlation.
The explicit results are given in   Eqs.   (\ref{increase1}, \ref{increase2}).
(2) the isotope effect of effective electron mass  Eq. (\ref{finalisotope}).
(3) the phonon spectral function computed in the fugacity expansion scheme,  Eq.  (\ref{phononresult}).

This paper is organized in the following way: In Sec. II we introduce AH model and present it in a form suitable for
semiclassical approximation.  In Sec. III we develop an instanton expansion of partition function and discuss its
renormalization group (RG) flow. 
In Sec. IV the phonon Green function is calculated in the fugacity expansion scheme.
We conclude this paper in Sec. V with discussion and summary. 
Some details of calculations can be found  in appendices.
%%%%%%%%%%%%%%%%%%%%%%%%%%%%%%%%%%%%%%%%%%%%%%%%%%%%%%%%%%%%%%%%%%%%%%%%%%%%%%%%%%%%%%%
\section{\label{sec2} Formulation}
The AH model is defined by the following Hamiltonian\cite{hewson}
\be
{\cal H}_{AH}={\cal H}_{el}+{\cal H}_{ph}+{\cal H}_{el-ph}.
\ee
\ba
{\cal H}_{el}&=&\sum_{k,\sigma} \epsilon_k \,c^\dag_{k\sigma} c_{k \sigma} 
+ \epsilon_f \sum_\sigma f^\dag_\sigma f_\sigma + U f^\dag_\uparrow f_\uparrow  f^\dag_\downarrow f_\downarrow \nonumber \\
&+& \sum_{k \sigma} \frac{1}{\sqrt{N_{{\rm lat}}}} ( V_k f^\dag_\sigma c_{k \sigma}+ V_k^*c^\dag_{k \sigma} f_\sigma).
\ea
\be 
{\cal H}_{ph}= \frac{1}{2}\Big( M \Omega^2 Q^2+ \frac{P^2}{M} \Big).
\ee
\be
{\cal H}_{el-ph}= g Q (  \sum_\sigma f^\dag_\sigma f_\sigma -1).
\ee
$\sigma=\uparrow, \downarrow$ is the spin index, and  $V_k$ is the hybridization matrix element.
 $f_\sigma$ is the impurity electron operator, and
$c_{ k  \sigma}$ is the conduction electron operator. $Q$ is local phonon coordinate.
$P$ is the conjugate momentum of $Q$ satisfying $[Q, P]= i \hbar $.
$M$ is ion mass and $\Omega$ is the oscillator frequency of dispersionless (Einstein) phonons.
$N_{{\rm lat}}$ is the number of lattice sites for the conduction electrons.
We will consider symmetric AH model specified by the relation $\epsilon_f + U/2=0$.

The conduction electrons $c_k$ can be integrated out exactly, yielding local impurity action.
\ba
& &S_{{\rm imp}}=-T \sum_{i \epsilon,\sigma} f^\dag_\sigma \, G_0^{-1}(i\epsilon) \,f_\sigma +
\int_0^\beta d\tau\, U f^\dag_\uparrow f_\uparrow  f^\dag_\downarrow f_\downarrow \nonumber \\
& &+\int_0^{\beta} d \tau \Big[  \frac{M}{2} (\partial_\tau Q)^2+\frac{M \Omega^2}{2} Q^2 
+g Q ( n-1 )\Big],
\ea
where $n_\sigma=f^\dag_\sigma f_\sigma,\;  n=n_\uparrow+n_\downarrow$, and
\ba
 G_0^{-1}(i\epsilon)&=&i \epsilon - \epsilon_f -\Sigma(i\epsilon), \nonumber \\
 \Sigma(i\epsilon)&=&\frac{1}{N_{{\rm lat}}}\,\sum_k \,\frac{|V_k|^2}{i\epsilon + \mu -\epsilon_k }.
\ea
For the hybridization matrix elements weakly dependent on momentum near Fermi surface,  
$ G_0^{-1}(i\epsilon)$ can be approximated by
\be
 G_0^{-1}(i\epsilon) = i \epsilon - \epsilon_f + i \Delta \, {\rm sgn}(\epsilon),
 \ee
 where $\Delta$ is the  hybridization enery scale 
\be
\Delta = \pi \langle  |V_k|^2 \delta(\omega-\epsilon_k) \rangle_{{\rm avg}}.
\ee
The average is done on the Fermi surface of conduction electron. 
We are interested in the strongly correlated regime $U \gg \Delta$.
The charge fluctuations at energy scale of order $U \gg  \Delta $ can be neglected in considering low
energy physics.
To separate out the charge fluctuations at high energy
 it is convenient to introduce the
Hubbard-Stratonovich (HS) transformation for the Hubbard $U$ interaction.
Two HS fields $\phi_c$ and $\phi_s$ are introduced to decouple on-site repulsion $U$ into
charge and spin channel, respectively.
\ba
& &e^{-U n_\uparrow n_\downarrow}=e^{-\frac{U}{4}(n^2-S_z^2)} \nonumber \\
& &=\int D[\phi_c,\phi_s]\,\exp[-\frac{\phi_c^2}{U}-\frac{\phi_s^2}{U} + i \phi_c n 
+ \phi_s S_z].
\ea
$S_z=n_\uparrow-n_\downarrow$ is the z-component of impurity spin.
Then, we choose the mean value of charge HS fields which can be absorbed into
$G_0^{-1}(i \epsilon)$ [cancelling $-\epsilon_f$]  and neglect its  high energy fluctuations.\cite{hamann}
Then electron-electron  interaction  [valid only at low energy] acts only in the spin channel, while 
phonon couples only in the charge channel.
Consequently at the leading approximation electron-electron interaction and  
electron-phonon interaction seem to decouple, however the interplay can be 
seen from the second order in expansion with respect to phonon fields [see below].

The resulting local action in imaginary time is 
\ba
\label{local}
S_{{\rm imp}}&=&-\int d \tau d \tau^\prime  \sum_\sigma  f^\dag_\sigma (\tau^\prime)   
G_0^{-1}(\tau^\prime-\tau)  f_\sigma (\tau) \nonumber \\
&+&\int d \tau \Big[ \frac{1}{U}\phi_{s}^2-\phi_{s} ( n_{ \uparrow} - n_{ \downarrow} )
 +g Q (n_\uparrow + n_\downarrow   -1) \nonumber \\
&+& \frac{M}{2} (\partial_\tau Q)^2+\frac{M \Omega^2}{2} Q^2 \Big].
\ea
The bare Green function $G_0(\tau^\prime-\tau)$ is given by
\be
G_0(\tau^\prime-\tau)= T \sum_{i \omega_n}\, 
\frac{e^{-i \omega_n (\tau^{\prime}-\tau)}}{i \omega_n + i \Delta {\rm sgn} \omega_n}.
\ee
The impurity electron $f$  can be integrated out exactly \cite{hamann}
for the action of the type of  Eq.  (\ref{local}).
\be
Z=\int D[ f, \phi_s, Q] \,e^{-S_{{\rm imp}}}=\int D[\phi_s,Q ]\,e^{-S[\phi_s,Q]},
\ee
where $S=S_P+S_K$ is given by 
\ba
\label{potential}
S_{P}&=&\int_0^\beta d \tau\Big[ \frac{\phi_s^2}{U}+\frac{M}{2}\Omega^2 Q^2
 \nonumber \\
&+&\sum _\sigma  \frac{\Delta}{\pi} \,
\big(-\zeta_\sigma\tan^{-1} \zeta_\sigma+\frac{1}{2}\,
\ln[1+\zeta_\sigma^2] \big ) \Big],
\ea
\ba
\label{kinetic}
S_{K}&=&\int_0^\beta d \tau \frac{M}{2} [\partial_\tau Q]^2
+\sum_\sigma  \frac{1}{2\pi^2} P \int_0^\beta \frac{d \tau d \tau^\prime}{\tau-\tau^\prime}\,
\zeta_\sigma(\tau)\, \nonumber \\
&\times &
\frac{d \zeta_\sigma(\tau^\prime)}{d \tau^\prime}\,
\frac{1}{\zeta^2_\sigma(\tau)-\zeta^2_\sigma(\tau^\prime)}\,
\ln \frac{1+\zeta_\sigma^2(\tau)}{1+\zeta^2_\sigma(\tau^\prime)},
\ea
where  $\zeta_\sigma(\tau)=[-\sigma \phi_s(\tau)+ g Q(\tau)]/\Delta$. $S_P$ and $S_K$ is the potential and the kinetic term, respectively.
It is convenient to  define an energy scale $E_L$ associated with    lattice relaxation (or equivalently polaron energy). 
\be
E_L=\frac{  g^2}{2 M \Omega^2}.
\ee
Following Ref.[\onlinecite{deppeler1,deppeler2}] we  define  dimensionless variables: 
\be
X= \frac{\phi_s}{\Delta},\;\;Y=\frac{g Q}{\Delta},\;\;
u=\frac{U}{\Delta},\;\;\lambda=\frac{E_L}{\Delta}, \;\; \gamma=\frac{\Omega}{\Delta}.
\ee

To develop semiclassical approximation we first need to   locate the classical minima of potential  term $S_P$.
When $u \gg \lambda$ and $u > \pi$, $S_P$ has two degenerate minima  $(X^*, Y^*)=(\pm  \eta_0, 0)$,
where $\eta_0 \sim   u/2 $.  We also assume $\gamma < 1$, 
even though the adiabatic expansion is not systematically employed in our study.
Our main interest resides in the case with strong electron correlation and weak electron-phonon interaction.
Accordingly, we  restrict our attention  to the regime
$u \gg 1 \gg \lambda$, so that the phonon part can be treated perturbatively.

There can exist other minima in the regime
$\lambda \gg u$.  This regime is essentially similar to the original Holstein model and has been studied in Ref.[\onlinecite{zeyher}]
in the semiclassical approximation scheme. Since we have treated the charge HS field $\phi_c$, which couples to
charge density like phonons,  only in a mean-field approximation, 
our present approach does not apply to the regime $\lambda \gg u$. Furthermore, the polaron state which is likely to 
occur in the regime $\lambda \gg u$ cannot be
described within  DMFT scheme with Lorentzian DoS. 

The quantum tunnelings between two minima $(X^*, Y^*)=(\pm  \eta_0, 0)$  are nothing but the spin flip processes in the
path integral framework.\cite{hamann, AGH}
The tunneling amplitude is determined by the kinetic term $S_K$.  
The key is to understand how the presence of weak electron-phonon interaction influences the spin flip processes
which are responsible for strongly correlated states.
To this end we integrate out the phonon field $Y$ in the second order.  There are no parts linear in $Y$. They vanish
upon the spin summation.  This integration over phonon field  is justified 
since the phonon field has a trivial minimum at origin, so that  we can 
employ harmonic approximation at low energy.
Concretely, we have to expand the action Eqs.  (\ref{potential}, \ref{kinetic}) as a power series in $Y$ up to the seocnd order.
The action $S=S_P+S_K$ expanded in $Y$ can be written as
\be
\label{start}
S \approx S_X + S_Y +S_{XY}.
\ee
$S_X$ is just the action Eqs.  (\ref{potential}, \ref{kinetic}) with $Q=0$, and it  is the starting point of the
Hamann's original anaylsis.\cite{hamann}
\ba
& &S_X=\frac{P}{\pi^2}\int_0^\beta \frac{d \tau d \tau^\prime}{\tau-\tau^\prime} X(\tau) \frac{d X(\tau^\prime)}{d \tau^\prime}\,
\frac{\ln \frac{1+X^2(\tau)}{1+X^2(\tau^\prime)}}{X^2(\tau)-X^2(\tau^\prime)} \nonumber \\
& &+\int_0^\beta d \tau \Big[  \frac{X^2}{u}+\frac{2}{\pi} \big( -X \tan^{-1}X+\frac{1}{2} \ln[1+X^2] \big) \Big]. 
\ea
$S_Y$ is the phonon part of Eq.  (\ref{local}) written in terms of $Y$.
\be
S_Y=\int_0^\beta d \tau \Big[ 
\frac{1}{4 \Delta \lambda \gamma^2} \left(\frac{d Y}{d \tau} \right)^2+
\frac{\Delta}{4\lambda} Y^2(\tau) \Big].
\ee
$S_{XY}$ embodies the interplay of spin fluctuation and the electron-phonon interaction, and has very complicated structure. 
The explicit expressions of the terms of $S_{XY}$ are listed in Appendix A.
A typical term of $S_{XY}$ looks as follows:
\ba
S_{XY,2}&=&-\frac{1}{\pi^2}\int d \tau d \tau^\prime \ln|\tau-\tau^\prime|
\,\dot{Y}(\tau) \dot{Y}(\tau^\prime)\,\nonumber \\
&\times&\Big( \int_0^1 d \xi \,\frac{1}{1+\xi X^2(\tau)}\frac{1}{1+\xi X^2(\tau^\prime)} 
\Big) \nonumber \\
&+&\frac{1}{\pi^2}\int d \tau d \tau^\prime \ln|\tau-\tau^\prime|\,
\dot{Y}(\tau) \dot{Y}(\tau^\prime)\, \nonumber \\
&\times&\Big(\int_0^1 d \xi \frac{2 \xi X^2(\tau)}{(1+\xi X^2(\tau))^2 
(1+ \xi X^2(\tau^\prime))}\Big ).
\ea

%%%%%%%%%%%%%%%%%%%%%%%%%%%%%%%%%%%%%%%%%%%%%%%%%%%%%%%%%%%%%%%%%%%%%%%%%
\section{\label{sec3} Renormalization Group Analysis}
The partition function can be computed by summing over instanton trajectories [ or hopping paths] in semiclassical approximation.
\cite{hamann}
The instanton trajectories are characterized by the hopping  locations $\{ t_i \}$. For simplicity we choose the same trajectories as
those of Hamann's, so that the derivative of $X$ is strictly zero unless $t_i < \tau < t_{i}+\tau_0$.
$\tau_0$ is the range in imaginary time where the hopping occurs.
We then expand the partition function $Z$ in the number of hoppings $N=0,2,4,\cdots$. 
The number of the hoppings is even due to the 
periodic boundary condition in imaginary time.
\be
Z=Z_{N=0}+Z_{N=2}+Z_{N=4}+\cdots.
\ee
It is convenient to normalize the partition function with respect to $Z_{N=0} \equiv Z_0$.
$Z_0$ is the partition function for the classical minima  $X=\pm \eta_0$. Since the action is an even functional in $X$ 
one can choose either minimum in computing $Z_0$. We choose $X= +\eta_0$.
\ba
\label{zeroth}
Z_0&=&\int D[Y] e^{-S_{{\rm ph}}[Y]-S_{X}[X= \eta_0]}, \nonumber \\
S_{X}[X= \eta_0]&=&\beta \Big[ \frac{\eta_0^2}{u}+\frac{2}{\pi} \big( - \eta_0 \tan^{-1}\eta_0  \nonumber \\
&+&\frac{\ln(1+\eta_0^2)}{2} \big) \Big], \nonumber \\
S_{{\rm ph}}[Y]&=&S_Y + S_{XY}[X=\eta_0] \nonumber \\
&=&\sum_\omega \Big[\frac{1}{4} \frac{1}{\Delta \lambda \gamma^2}
|\omega|^2 +\frac{1}{\pi (1+\eta_0^2)^2} |\omega|  \nonumber \\
&+& \frac{\Delta}{4 \lambda}(1-\frac{\lambda}{\pi(1+\eta_0^2)} ) \Big ]
|Y(i\omega)|^2,
\ea
Note two new terms coming from $ S_{XY}[X=\eta_0]  $ in $S_{{\rm ph}}[Y]$. 
From the action $S_{{\rm ph}}$ in Eq.  (\ref{zeroth}) the phonon propagator in the absence of instantons is easily obtained.
\be
\label{phononwokink}
\langle Y(i\omega)  Y(-i \omega) \rangle_{N=0} = ( \frac{g}{\Delta})^2
 \frac{M^{-1}}{|\omega|^2+ E^* |\omega|+\omega_0^2},
\ee 
where [ the condition $\eta_0 \gg 1$ is used]
\ba
\label{newdef}
\omega_0^2 &\sim & \Omega^2  \Big(1- \frac{\lambda}{ \pi \eta_0^2} \Big), \nonumber \\ 
E^* &\sim& E_L \frac{4}{\pi}\, \frac{\gamma^2}{\eta_0^4}  \ll E_L, \nonumber \\ 
 \alpha &\equiv& E^*/\Delta.
\ea
Note that the frequency of phonon decreases as the electron phonon coupling $g$ increases
in the strong coupling limit of large $U$. This agrees with the result of Ref.[\onlinecite{jpc}].
 $E^*$ is a new energy scale involving both electron-phonon and electron-electron 
interaction.
Note that $E^*/\Omega = \lambda \gamma/ \eta_0^4 \ll 1$.

Being  expanded in the number of instantons the normalized partition function can be expressed as
 \ba
 \label{instantongas}
\frac{Z}{Z_0}&=&1+\sum_{n=1}  (z \tau_0)^{2n}\, \int_0^{\beta}\,\frac{d t_{2n}}{\tau_0} \,
\int_0^{t_{2n}-\tau_0}\,\frac{d t_{2n-1}}{\tau_0}\cdots
 \nonumber \\
&\times&\,\int_0^{t_{2}-\tau_0}\,\frac{d t_{1}}{\tau_0}\exp[+V(t_1,\cdots,t_{2n})],
\ea
where $V(\{t_i\})$ is the energy of instanton gas which includes the corrections induced by phonon coupling. 
$z$ is the fugacity of instanton gas.
$Z_0$ provides a denominator for phonon averaging, so that  the explicit form of 
$\langle Y(i\omega)  Y(-i \omega) \rangle_{N=0}$ can be used in the evaluation of partition function.
The integration over phonon field $Y$  induces a few 
corrections to the original Coulomb (kink) gas action \cite{AGH,hamann}of 
Kondo problem without phonons [$K_2$  term of Eq. (\ref{energy})] . 
These corrections represent the interplay of electron correlation and 
electron-phonon interaction.
\ba
\label{energy}
V(t_1,\cdots,t_{2n})&=&K_1 \sum_{i > j}  (-)^{i-j}\,\ln \left(\frac{t_i-t_j}{\tau_0} \right)
\nonumber \\
&+&K_2  \sum_{i > j} \,\ln \left( \frac{t_i-t_j}{\tau_0} \right )\,d(t_i,t_j).
\ea
\be
d(t_i,t_j)=\frac{1}{\omega_0^2}\,\langle \frac{d Y}{d \tau}(\tau=t_i) \frac{d Y}{d \tau}(\tau=t_j) \rangle.
\ee
The intial (bare) values  of coupling constants of Eq.  (\ref{energy}) are 
\ba
\label{k1}
K_1 & \sim & 2 \left( \frac{2 \tan^{-1} \eta_0}{\pi} \right )^2 + c_1 \frac{\bar{\lambda} {\bar{\gamma}}}{\eta_0^2},
  \\
 \label{k2}
K_2 & \sim & -\frac{ (\omega_0 /\Delta)^2}{4 \eta_0^2},  \\
\label{fugacity}
z \tau_0 & \sim & \exp\Big[-\ln \eta_0 +  \frac{\bar{\lambda}\bar{ \gamma}}{2 \eta_0} \Big ].
\ea
$c_1$ is a numerical constant of order unity and 
\be
\bar{ \gamma} = \omega_0/\Delta, \;\; \bar{\lambda}=\frac{g^2}{2 M \omega_0^2 \Delta}.
\ee
The corrections in Eqs.  (\ref{k1},\ref{fugacity}) [ $\frac{\bar{\lambda}\bar{ \gamma}}{2 \eta_0}$ ]
reveal the expansion parameters as noted by Deppeler and Millis.\cite{deppeler1}
The influences of phonons are reflected in the $K_2$ term of the energy $V(\{t_i \})$ and in the change of 
initial values from those without phonons.
The most drastic difference of $K_2$ term from $K_1$ is the absence of the alternating factor $(-1)^{i-j}$.
This  absence  of the alternating factor  indirectly demonstrate the fact that the phonon interacts with electron via
 charge channel not spin channel.

The renormalization group equation (RGE) of the instanton gas Eq.  (\ref{instantongas})
  can be derived by following the procedure of  Ref. [\onlinecite{AGH}]. 
For details, see appendix II.
\ba
\label{fuga}
\frac{d x}{d l}&=& x \big[ (1-\frac{K_1}{2}) + 2 x^2  K_2 g_2 \big],\\
\label{knormal}
\frac{d K_1}{d l}&=& -4 K_1 x^2,\;\;\frac{d K_2}{dl}=0,
\ea
where $dl= d \tau_0 /\tau_0,  \;\; x= z \tau_0$. $g_2$ is defined in Eq.  (\ref{g2}).
At the present level of approximation the influence of phonon coupling is manifest only in the renormalization of
the fugacity $x$. 
For the sake of the comparison with the original Kondo model it is convenient to
introduce a new variable $y=2-K_1$.  The initial value of $y$ is smaller than 1 as can be seen Eq.  (\ref{k1}).
Using the result Eq.  (\ref{g2result}), the RG Eqs.  (\ref{fuga},\ref{knormal}) can be re-expressed as
\ba
\label{fugarg}
\frac{d x }{dl}&=&\frac{x}{2}\Big[y + x^2 \alpha  e^{-\omega_0 \tau_0} \Big],\nonumber \\
\frac{d y}{d l}&=& 8 x^2 - 4 y x^2 \sim 8 x^2,\;\; \frac{d K_2}{ d l}=0.
\ea
$\alpha$ is defined in Eq. (\ref{newdef}).
Since the phonons have been treated perturbatively from the outset
we are not far from the original Kondo physics which does not have phonons. 
Thus, even if  the SU(2) spin symmetry is not manifest in our approximate RGE, 
we can trace  the isotropic ray $y=4x$ of Eq.  (\ref{fugarg}) in approaching strong coupling regime.
Due to the presence of explicit energy scale $\omega_0$ in Eq.  (\ref{fugarg}), 
 {\it two} energy scales $T^{(0)}_K$,  $\omega_0$ should be taken into account     in the analysis of Eq.  (\ref{fugarg}).
$T^{(0)}_K \sim  D e^{- U /\Delta}$ is the Kondo temperature in the absence of phonons. $D$ is an energy cutoff.

First consider the case of $ \omega_0  \ll T_K^{(0)}$.
Then the scaling should stop near $l_K \sim  \ln \frac{D}{T_K^{(0)}}  <   \ln \frac{D}{\omega_0 }=l_{\omega_0}$.
This is because  beyond $l_K$ the RG flow begins to enter the strong coupling regime where our perturbative 
scaling equations do not apply.
In this region of $l < l_K $,  $e^{-\omega_0 \tau_0} \sim 1$ holds.
Integrating  RG equation until $x$ becomes order of unity we obtain
\ba
\label{newkondo1}
T_K &\sim& D \exp[-\frac{1}{2 x(0)}]\times \exp[ \frac{\alpha}{8}\, \ln(\frac{1}{x(0)})] \nonumber \\
& \sim &T_K^{(0)}e^{ +\frac{\lambda}{2 \eta_0^2}+\frac{\alpha \ln(\eta_0)}{8}}.
\ea
According to the result Eq.  (\ref{newkondo1}) the electron-phonon interaction {\it increases} the Kondo energy scale, and 
this is in agreement with the results of NRG studies.\cite{hewson,jpc}
The  amount of increase  is given by
\be
\label{increase1}
\delta T_K \sim (\frac{\lambda}{2 \eta_0^2}+  \frac{\alpha}{8}\,\ln \eta_0)\,T_K^{(0)}.
\ee
Second consider the case of $ T_K^{(0)} \ll \omega_0 $.
In this case,  the RG flow evolves until $l=l_{\omega_0}= \ln \frac{D}{\omega_0 }$, 
and  then the second term of $\frac{dx}{dl}$ in Eq.  (\ref{fugarg}) becomes
completely negligible beyond $l_{\omega_0}$.
In the region $l > l_{\omega_0}$,  $x(l_{\omega_0})$ plays a role of   a new initial value.
Integration RG equation starting from $l=l_{\omega_0}$ until $x$ becomes the order of unity we obtain 
the Kondo temperature for the second case.
\be
T_K  \sim  \omega_0 e^{-1/2 x(l_{\omega_0})}.
\ee
The direct evaluation of $ x(l_{\omega_0})$ gives 
\be
\label{newkondo2}
T_K \sim T_K^{(0)}\exp\Big[\frac{\lambda}{2 \eta_0^2}+\frac{\alpha}{2 \eta_0} \ln \frac{1}{\gamma_0} \Big],\;
\gamma_0=\omega_0/\Delta.
\ee
The increase of Kondo temperature is given by
\be
\label{increase2}
\delta T_K  \sim \Big[\frac{\lambda}{2 \eta_0^2}+\frac{\alpha}{2 \eta_0} \ln \frac{1}{\gamma_0} \Big]  T_K^{(0)}.
\ee 
Note that the results Eqs. (\ref{newkondo1},\ref{newkondo2}) are obtained in the regime of strong electron correlation where
ME theory does not apply.
Both results  Eqs. (\ref{newkondo1},\ref{newkondo2})  show the electron-phonon interaction increases  the Kondo temperature but
by rather small amount. The increase of  Kondo temperture could be heuristically understood in  somewhat unphysical 
antiadiabatic limit $\Omega \gg  \Delta$ [see Eq. (\ref{effectiveU}, \ref{kondoeffective})].
 This point will be elaborated  further in sec V.  
In the antiadiabatic limit the retardation effects of phonons are negligible.
Our results of Kondo temperature are valid in the strong electron correlation 
and fully incorporate the retardation effects of phonons.

The results, Eqs. (\ref{newkondo1},\ref{newkondo2}), can be interpreted in the light of lattice system within the context of 
DMFT.
The Kondo singlet state of impurity problem corresponds to the Fermi liquid in the original (large-{\it d}) lattice
system (see the Sec. VII of Ref. [\onlinecite{dmft}]).  
Here the momentum dependence of the electron self-energy is neglected, and the effect of correlation is encapsulated in the
wave-function renormalization $Z$:
\be
Z(i\omega)= \Big[ 1- \frac{\partial \Sigma(z)}{\partial z} \Big ]_{z \to i\omega},
\ee
which gives the residue of the quasiparticle. $Z$ is related to the mass enhancements of quasiparticles as
\be
Z^{-1}=m^{*}/m_b,
\ee
where $m^{*}$ is the effective mass, and $m_b$ is the bare mass given by band theory.
The Kondo temperature itself can be identified with the {\it effective} bandwidth $B_{{\rm eff}}$ of lattice system.
Then the mass enhancement can be estimated to be
\be
\label{mass}
Z^{-1}=m^{*}/m_b \sim  D/ T_K.
\ee
$D$ is the order of half bandwidth. In the DMFT with Lorentzian Dos it can be identified with $\Delta$.
The above estimate follows from the observation 
$B_{{\rm eff}} \sim  p_F^2/ m^*, \; D\sim p_F^2/m$. $p_F$ is Fermi momentum.
Combining  Eq.  (\ref{mass}) with our results Eqs. (\ref{newkondo1},\ref{newkondo2}), the issue of isotope effect on 
the mass enhancement can be addressed.
The ratio of mass enhancements in the presence and absence of phonons is given by
\be
\label{isotope}
\frac{m^{*}/m_b}{m^{*,(0)}/m_b} \sim  T_K^{(0)}/T_K \sim 1 - \delta T_K /  T_K^{(0)},
\ee
where $m^{*,(0)}$ is the effective mass in the absence of phonons.
Thus, the isotope effect is {\it negative}  in that the electron-phonon interaction which depends on 
ion masses  {\it decreases} the effective mass of electrons.
This isotope effect manifests itself in the specific heat coefficient, the $T^2$ coefficient of low temperature resistivity,
the Drude weight of optical conductivity, 
and in other physical quantities which depend on the effective bandwith $B_{{\rm eff}}$ in a crucial way.
Our results show that the isotope effect is  very small in the regime of strong electron correlation.
\be
\label{finalisotope}
\frac{m^{*}/m_b}{m^{*,(0)}/m_b} \sim 1  -  E_L  \Delta  / U^2.
\ee
The proof of Eq. (\ref{finalisotope}) is beyond the validity of ME theory, however.
 %%%%%%%%%%%%%%%%%%%%%%%%%%%%%%%%%%%%%%%%%%%%%%%%%%%%%%%%%%%%%%%%%%%%%%%%%%%%%%%%%%%%%%%%%%%%%%%%%%%%%%%%%%%%%%%
\section{Phonon correlation function}
The phonon correlation function is defined by
\be
\label{corr}
D(t-t^\prime)=\frac{\int D[X,Y] \, e^{-S} \, Y(t) Y(t^\prime)}{\int D[X,Y] \, e^{-S} }.
\ee
The correlation function of original phonon coordinate $Q$ is related with Eq.  (\ref{corr}) by
\be
\langle Q(t) Q(t^\prime) \rangle=(\frac{\Delta}{g})^2 D(t-t^\prime).
\ee
The denominator of the right hand side of Eq.  (\ref{corr}) is just the partition function stuied in the 
previous section  and will be written as $Z=Z_0 (Z/Z_0)$.

We will compute the phonon correlation function in the framework of fugacity expansion.
The phonon correlation function in the absence of instanton is given by Eq.  (\ref{phononwokink}).
As has been discussed in Sec. \ref{sec3} the fugacity is relevant at low energy, thus our
result for the phonon correlation is expected to be valid only in the  high frequency/temperature
regime $( |\omega|,  T)  > T_K$. 

The first nontrivial contribution comes from the  trajectories with an instanton-antiinstanton pair.
We rewrite Eq.  (\ref{corr}) as
\be
\label{corr1}
D(t-t^\prime)=\frac{ Z_{{\rm ph}}^{-1} \int D[X,Y]  e^{-S+S_X[\eta_0]}  Y(t) Y(t^\prime)}{Z/Z_0},
\ee
where $Z_{{\rm ph}}=\int D[Y] e^{-S_{{\rm ph}}[Y]}$ [see Eq. (\ref{zeroth})].
Let us write the  exponent of Eq.  (\ref{corr1}) in the form
\be
-S+S_X[\eta_0]=-S_{{\rm ph}}-\delta S_X-\delta S_{XY},
\ee
where 
\ba
\delta S_X&=& S_X[X] - S_X[X=\eta_0], \nonumber \\
\delta S_{XY} &=& S_{XY}[X,Y]-  S_{XY}[X=\eta_0,Y].
\ea
By construction $\delta S_X$ and $\delta S_{XY}$ vanish in the absence of instantons.
Now the structure of fugacity expansion is manifest in the form of Eq.  (\ref{corr1}).
\be
D(t-t^\prime)=D_0(t-t^\prime)+ D_2(t-t^\prime) + \cdots,
\ee
where $D_n$ is proportional to $z^n$.
Since we are only interested in the connected diagrams we have to select the connected components
among contributions to $D_2(t-t^\prime)$. 

The structure of $\delta S_{XY}$ is very complicated, and a typical term ( which, in fact, will give the  most dominant
contribution ) looks like
\ba
& &\delta S_{XY,2}=-\frac{1}{\pi^2}\int d \tau d \tau^\prime \ln|\tau-\tau^\prime|
\,\dot{Y}(\tau) \dot{Y}(\tau^\prime)\, \int_0^1 d \xi\nonumber \\
& & \times \Big[\frac{1}{1+\xi X^2(\tau)}\frac{1}{1+\xi X^2(\tau^\prime)} 
-\frac{1}{(1+\xi \eta_0^2)^2}\Big ] \nonumber \\
& &+\frac{1}{\pi^2}\int d \tau d \tau^\prime \ln|\tau-\tau^\prime|\,
\dot{Y}(\tau) \dot{Y}(\tau^\prime)\,\int_0^1 d \xi \nonumber \\
& & \times\Big[\frac{2 \xi X^2(\tau)}{(1+\xi X^2(\tau))^2 
(1+ \xi X^2(\tau^\prime))}-\frac{2 \xi \eta_0^2}{(1+\xi \eta_0^2)^3} \Big ].
\ea 
Next we need to evaluate the action $\delta S_X$ and $\delta S_{XY}$ 
in the subspace of two instantons which are parameterized by their locations $t_1,t_2$.
Then the connected component of  $D_2(\tau-\tau^\prime)$ can be written
\ba
& &D_2(t-t^\prime)= \frac{z^2}{Z_{{\rm ph}}}\,\int D[Y] e^{-S_{\rm{ph}}}\, 
\int_0^\beta d t_2  \int_0^{t_2-\tau_0}\,d t_1\nonumber \\
&\times& \,e^{-K_1 \ln \frac{t_2-t_1}{\tau_0}}\,
e^{-\delta S_{XY}(t_1,t_2)}\,Y(t) Y(t^\prime).
\ea
We will treat $\delta S_{XY}(t_1,t_2)$ perturbatively as in the evaluation of the partition function.
In the expansion $e^{-\delta S_{XY}}=1-\delta S_{XY} + \cdots$, the first term cancels the disconnected piece coming from the 
denominator $Z/Z_0$.
Among various contributions to $D_2(t-t^\prime)$ the term containing two time derivatives of  $Y$, $\delta S_{XY,2}$,
gives the most dominant one.
The calculation of $D_2(t-t^\prime)$ can be cast into the following form  
\ba
& &D_2(t-t^\prime)=D_0(t-t^\prime) +\int d t_1 d t_2 \nonumber \\
& &\times D_0(t-t_1) \Sigma(t_1-t_2)  D_0(t_2-t^\prime).
\ea
The function $\Sigma(t_1-t_2)$ plays a role of self-energy.
The explicit form of  self-energy is
\be
\Sigma(t_1-t_2)  \sim  z^2 \frac{ \omega_0^2}{\eta_0^4}\, 
\frac{\ln(\frac{|t_1-t_2|}{\tau_0})}{(\frac{|t_1-t_2|}{\tau_0})^{K_1}}.
\ee
In frequency space it becomes  
\be
\Sigma(i\omega) \sim  z^2 \frac{ \omega_0^2}{\eta_0^4}\,  |\omega|^{K_1-1}  \ln \frac{1}{|\omega|}.
\ee
Thus the phonon correlation function which is accurate up to the second order of $z^2$ is
\be
\label{phononresult}
D(i\omega)=\frac{2 \lambda \gamma^2 \Delta}{
\omega^2+ E^* |\omega|  + \omega_0^2+ z^2 \gamma^2 E^*  |\omega|^{1-y} \ln \frac{1}{|\omega|}}.
\ee
Apparently the  self-energy term behaves very singular   at low energy, but we have to keep in mind that the 
Eq.  (\ref{phononresult}) was derived in the high temperature/frequency regime $(|\omega|, T) > T_K$. 

Let us compare our results with those by NRG.
The NRG calculation of phonon Green function by Hewson and Meyer has been improved in Ref.[\onlinecite{jpc}].
The imaginary part of phonon Green function in the regime of large $U$ has the following features:
(1) the decrease of phonon frequency with increasing phonon coupling
(2)  broadended phonon peak (3) absence of peak near zero frequency.
These feature are consistent with our results.
First the renormalized phonon frequency 
\be
\omega_0  \sim \Omega \Big(1- \frac{2}{\pi} \frac{E_L \Delta}{U^2} \Big).
\ee
clearly shows the decrease of phonon frequncey with increasing $\lambda$.

This result can be compared with that derived in the {\it weak} coupling  RPA type perturbative calculation.\cite{jpc}
\be
\omega_0 \sim  \Omega \Big(1-  \frac{E_L}{ \pi \Delta} \Big).
\ee
Second the renormalization of the phonon propagator can be reliably discussed if the interested 
energy  scale is {\it larger} than Kondo temperature.
If $\omega_0  \agt T_K$ our result for the phonon Green's function can be applied 
in the neighborhood of $\omega_0$.
The broadening is basically determined by  the energy scale $E^*$  which increases with  increasing 
electron-phonon coupling. This feature is also consistent with the NRG result.
In the perturbative RPA approach the broadening width is predicted to be linear in $\omega$ while
our results predict sublinear dependence of width $\omega^{1-y}$ which is stronger at low energy.
Notice that  the anomalous self-energy term represents strongly \textit{temperature dependent} broadening
since it originates from the electronic processes happening near Fermi energy.
This feature can be checked in principle by NRG calculation at finite temperature. 
%%%%%%%%%%%%%%%%%%%%%%%%%%%%%%%%%%%%%%%%%%%%%%%%%%%
\section{Discussions and Summary}
-\textit{Antiadiabatic limit} Our results can be  compared with those obtained in the antiadiabatic limit.
In fact, the antiadiabatic limit is relevant to rather few materials, but 
theoretically it provides a convenient benchmark.
The antiadiabatic limit can be easily taken by integrating out phonons.
The integration results in an effective {\it retarded} and {\it attractive} interaction between electrons.
\ba
\label{retarded}
S_{ee,eff}&=&-\int d \tau d \tau^\prime\,\frac{g^2}{2M}\,
 \big[ T \sum_\omega \frac{e^{-i \omega (\tau-\tau^\prime)}}{\omega^2+\Omega^2} \big ]\nonumber \\
&\times& [n(\tau)-1]\,[n(\tau^\prime)-1].
\ea
If the phonon frequency $\Omega$ is larger than typical electronic energy scale such as 
 $\Delta$,
 then the frequency dependence in the denominator of $\frac{e^{-i \omega (\tau-\tau^\prime)}}{\omega^2+\Omega^2}$
can be neglected. The summation over frequency 
gives the delta function in  imaginary time,
and Eq.  (\ref{retarded}) becomes 
\be
S_{ee,eff}=-2 E_L \int d \tau \, n_\uparrow  n_\downarrow,\quad 
\Omega \gg \Delta.
\ee
Now $S_{ee,eff}$ can be combined with the  on-site Coulomb repulsion $U$ term, leading to
\be
\label{effectiveU}
U_{{\rm eff}}=U-2 E_L.
\ee
Note that this effective Anderson model  is valid only in the  limit $\Omega \gg \Delta$.
In our regime of interest  $ U \gg E_L$, $U_{{\rm eff}}$ is sufficiently large so that the physics in Kondo regime
applies.
One can expect a qualitative change of behavior if $U_{{\rm eff}}=U-2 E_L$ becomes negative.
This is indeed confirmed by NRG studies.\cite{hewson,jpc}
Our treatment is not applicable to those regimes, and the polaron physics describes the regime adequately.\cite{hewson,jpc}
The Kondo temperature of this effective Anderson model in the limit $U \gg E_L$  is given by\cite{thebook}
\be
\label{kondoeffective}
T_{K,{\rm eff}} \sim \sqrt{\frac{U_{{\rm eff}} \Delta}{2}}\,\exp[- \frac{\pi U_{{\rm eff}}}{8 \Delta}].
\ee
The change of  Kondo temperature due to electron-phonon interaction given by
\be
\label{changeeffective}
\delta T_{K, {\rm eff}} \sim T_K^{(0)} \Big(\frac{\pi}{4} \lambda -  \frac{\lambda}{u} \Big),
\ee
in the limit of small $\lambda$.
Eq.  (\ref{changeeffective}) appears to be much larger than  Eq.  (\ref{increase1}, \ref{increase2}).
This also demonstrates that the physics in our regime is essentially different from that of antiadiabatic limit.

-\textit{Adiabatic Expansion} Deppeler and Millis studied the problem of electron-phonon interactions in correlated systems using 
adiabatic expansion of DMFT without any specific ground state assumed.\cite{deppeler1,deppeler2}
They have obtained the phonon Green function
\be
D^{-1}(i\omega)=1- \frac{\lambda}{\lambda_c} + \omega^2 + \lambda \gamma \alpha_p |\omega| + o(\gamma^2).
\ee
The factor $1- \frac{\lambda}{\lambda_c}$ represents the softening of phonon frequency in our notations.
At half-filling and in the Hatree-Fock approximation, they obtained $\lambda_c  \propto  1/u^3$ while
our result is $\lambda_c  \propto   1/u^2$.  
This is not real contradiction since  our result does not apply to the half-filling case in the limit of large $U$.
They have also identified the renormalized expansion parameters 
\be
\bar{\gamma}=\gamma (1- \lambda/\lambda_c)^{1/2},\;\;
\bar{\lambda}=\frac{\lambda}{1-\lambda/\lambda_c},
\ee
which coincide with our results.
In our case,  basically we obtain such expansion parameter through
\be
\int \frac{d \omega}{2\pi} \langle Y(-i \omega) Y (+i\omega) \rangle_{N=0} =\frac{1}{2} \bar{\lambda} \bar{\gamma}.
\ee
In Ref.[\onlinecite{deppeler2}] the isotope effect on the effective electron mass has been investigated.
Away from half-filling and \textit{in the absence of strong electron correlation} with the semicircular DoS they obtained 
the \textit{positive} mass enhancement which increases with decreasing phonon frquency, and increases with
increasing $\bar{\lambda}$.
This should be also compared with our {\it negative} mass enhancement [see Eq.  (\ref{finalisotope})],
 where the strong electron correlation plays an essential role.

-\textit{Role of the orbital degrees of freedom}
It should be pointed out that our results depend very crucially on the detailed structure of 
electron-phonon coupling. We wish illustrate this point by an example from the DMFT study on CMR manganese oxides
in metallic regime, such as $\text{La}_{0.7} \text{Sr}_{0.3} \text{MnO}_3$.\cite{urushibara,nagaosa}
The local action of this system is
\ba
\label{multiorbital}
S_{{\rm imp}}&=&-\int d \tau d \tau^\prime  \sum_\sigma  f^\dag_\sigma (\tau^\prime)   
G_0^{-1}(\tau^\prime-\tau)  f_\sigma (\tau) \nonumber \\
&+&\int d \tau \Big[ \frac{1}{U}\phi_{s}^2-\phi_{s} ( n_{ \uparrow} - n_{ \downarrow} )
 +g Q (n_\uparrow - n_\downarrow  ) \nonumber \\
&+& \frac{M}{2} (\partial_\tau Q)^2+\frac{M \Omega^2}{2} Q^2 \Big].
\ea
In this system the real spin is fully polarized, and the index $\sigma$ denotes the \textit{orbital} degrees of freedom of $d$-band
$\uparrow=d_{x^2-y^2}, \downarrow=d_{3 z^2-r^2}$. Here $Q$ is  one of the Jahn-Teller phonons (the third component). 
In spite of the similarity of  the local action Eq.  (\ref{multiorbital}) with  the Eq.  (\ref{local})  
\textit{except for} the difference in the electron-phonon coupling,  the outcomes are drastically different.
For the model  Eq.  (\ref{multiorbital}), 
the phonon frequency is \textit{hardened}  $\omega_0 = \Omega  \sqrt{(U + 4 E_L)/U}$,
and Kondo temperature \textit{decreases} as follows:
\be
\label{orbitalkondo}
T_K \sim \Delta  \exp[- R \frac{ U+ 4 E_L}{\Delta}],\;\; R \gg 1.
\ee 
The reduction factor $R$ is related to the overlap of the phonon wave functions in the process of instanton tunneling.
In fact, major results are just opposite to ours.  This indicates that the orbital degrees of freedom being associated with
the diverse forms of  couplings with electrons exhibit rather rich physics.

-\textit{Summary}  We have studied AH model in the framework of semiclassical approximation, and 
interpreted the results in the light of DMFT. We obtained the analytic results for the Kondo temperature
which are renormalized by weak electron-phonon interaction and for the phonon Green function which are valid
at frequency higher than Kondo temperature. We have compared our results with the existing NRG results  and other related 
works, and they are consistent with each other in their common regime of validity.

%%%%%%%%%%%%%%%%%%%%%%%%%%%%%%%%%%%%%%%%%%%%%%%%%%%%%%%%%%%%%%%
\begin{acknowledgments}
H.C.L  was  supported by the Korea Science and Engineering
Foundation (KOSEF) through the grant No. R01-1999-000-00031-0 and  through
the Center for Strongly Correlated Materials Research (CSCMR) (2003).
H. Y. Choi was supported by the by the Korea Science and Engineering
Foundation (KOSEF) through the grant No. R01-1999-000-00031-0, the 
Ministry of Education through Brain Korea 21 SNU-SKKU Program, and CNNC.
\end{acknowledgments}
%%%%%%%%%%%%%%%%%%%%%%%%%%%%%%%%%%%%%%%%%%%%%%%%%%%%%%%%%%%%%%%%
\appendix
%%%%%%%%%%%%
\section{The detailed form of $S_{XY}$}
In this section we exhibit all the terms comprising $S_{XY}$.
The expansion of the action as a power series in $Y$ is facilitated
 by utilizing the following parametric integral
\ba
I_0&=&\int_0^1 \frac{d \xi}{[1+ \xi \zeta^2_\sigma(\tau)][1+ \xi \zeta^2_\sigma(\tau^\prime)]} \nonumber \\
&=&
\frac{1}{\zeta^2_\sigma(\tau)-\zeta^2_\sigma(\tau^\prime)}\,
\ln \frac{1+\zeta^2_\sigma(\tau)}{1+\zeta^2_\sigma(\tau^\prime)}.
\ea
$S_{XY}$ consists of 10 terms which are numbered as $S_{XY,i}, i=1,\ldots,10$.
Every term contains  {\it two} time derivatives except for $S_{XY,1}$.
\be
S_{XY,1}=-\frac{\Delta}{\pi}\,\int \,\frac{Y^2(\tau)}{1+X^2(\tau)}.
\ee
\ba
& & S_{XY,2}=-\frac{1}{\pi^2}\int d \tau d \tau^\prime \ln|\tau-\tau^\prime|
\,\dot{Y}(\tau) \dot{Y}(\tau^\prime)\,\nonumber \\
& & \times\Big( \int_0^1 d \xi \,[\frac{1}{1+\xi X^2(\tau)}\frac{1}{1+\xi X^2(\tau^\prime)} 
-\frac{1}{(1+\xi \eta_0^2)^2}]\Big) \nonumber \\
&&+\frac{1}{\pi^2}\int d \tau d \tau^\prime \ln|\tau-\tau^\prime|\,
\dot{Y}(\tau) \dot{Y}(\tau^\prime)\,\Big(\int_0^1 d \xi \nonumber \\
& & \times [\frac{2 \xi X^2(\tau)}{(1+\xi X^2(\tau))^2 
(1+ \xi X^2(\tau^\prime))} -\frac{2 \xi \eta_0^2}{(1+\xi \eta_0^2)^3}]\Big ).
\ea 
\ba
 S_{XY,3}&=&-\frac{1}{\pi^2}\int d \tau d \tau^\prime \ln |\tau-\tau^\prime|\,
\dot{X}(\tau) \dot{X}(\tau^\prime) \nonumber \\
&\times& \Big[ \int_0^1 d \xi \frac{ (3 X^2(\tau) \xi^2-\xi)}{[1+\xi X^2(\tau)]^3}\,
Y^2(\tau)\,\frac{1}{[1+\xi X^2(\tau^\prime)]}  \nonumber \\
&+&
\int_0^1 d \xi \frac{Y^2(\tau^\prime)}{[1+ \xi X^2(\tau)]}\,
 \frac{ (3 X^2(\tau^\prime) \xi^2-\xi)}{[1+\xi X^2(\tau^\prime)]^3} \nonumber \\
 &+&\int_0^1 d \xi \frac{ 4 \xi X(\tau) X(\tau^\prime) Y(\tau) Y(\tau^\prime)}{
 [1+\xi X^2(\tau)]^2 [1+\xi X^2(\tau^\prime)]^2} \Big ].
 \ea
\ba
& & S_{XY,4}=-\frac{1}{\pi^2}\int d \tau d \tau^\prime \ln |\tau-\tau^\prime|\,
\dot{X}(\tau^\prime) X(\tau) \nonumber \\
& &\times\Big[ \int_0^1 d \xi \frac{d }{d \tau}\left(
\frac{ (3 X^2(\tau) \xi^2-\xi)}{[1+\xi X^2(\tau)]^3}\right)\,
\frac{ Y^2(\tau)}{[1+ \xi X^2(\tau^\prime)]} \nonumber \\
& &+\int_0^1 d \xi \frac{d}{d \tau} \left( \frac{1}{1+\xi X^2(\tau)} \right)\,
\frac{ (3 X^2(\tau^\prime) \xi^2-\xi)Y^2(\tau^\prime)}{[1+\xi X^2(\tau^\prime)]^3} \nonumber \\
&+&\int_0^1 d \xi 4 \xi \frac{d}{d\tau}
\left( \frac{X(\tau)}{[1+\xi X^2(\tau)]^2} \right ) 
\frac{ X(\tau^\prime) Y(\tau) Y(\tau^\prime)}{[1+\xi X^2(\tau^\prime)]^2} \Big].
\ea
\ba
& &S_{XY,5}=\frac{1}{\pi^2}\int d \tau d \tau^\prime \ln |\tau-\tau^\prime|\,
\dot{X}(\tau^\prime)\,Y(\tau)\,\nonumber \\
&\times&\Big[ \int_0^1 d \xi 2 \xi \, \frac{d}{d \tau}\,
\left( \frac{X(\tau)}{[1+\xi X^2(\tau)]^2} \right ) \frac{Y(\tau)}{[1+\xi X^2(\tau^\prime)]}
\nonumber \\
&+&\int_0^1 d \xi \, 2 \xi\,\frac{d}{d \tau} \left( \frac{1}{1+ \xi X^2(\tau)} \right )
\,\frac{ X(\tau^\prime) Y(\tau^\prime)}{[1+ \xi X^2(\tau^\prime)]^2}\Big ].
\ea
\ba
S_{XY,6}&=&-\frac{1}{\pi^2}\,\int d \tau d \tau^\prime \ln|\tau-\tau^\prime|\,
\dot{X}(\tau^\prime)\,X(\tau)\,\nonumber \\
&\times& \Big[ 
\int_0^1 d \xi \,\frac{3 X^2(\tau) \xi^2-\xi}{[1+\xi X^2(\tau)]^3}\,
\frac{d Y^2(\tau)}{ d \tau}\,\frac{1}{1+\xi X^2(\tau^\prime)}  \nonumber \\
&+&
\int_0^1 d \xi 4 \xi^2\,
\frac{ X(\tau) X(\tau^\prime)\,\frac{d Y(\tau)}{d \tau} Y(\tau^\prime)}{
[1+\xi X^2(\tau)]^2 [1+\xi X^2(\tau^\prime)]^2 } \Big].
\ea
\ba
& &S_{XY,7}=-\frac{1}{\pi^2}\,\int d \tau d \tau^\prime\,
\ln|\tau-\tau^\prime|\,\dot{Y}(\tau^\prime)\,Y(\tau) \nonumber \\
&\times& \int_0^1 d \xi \, \frac{d }{d \tau} \Big[ \frac{1}{1+\xi X^2(\tau)} \Big]\,
\frac{1}{1+\xi X^2(\tau^\prime)}.
\ea
\ba
& &S_{XY,8}=+\frac{1}{\pi^2}\,\int d \tau d \tau^\prime \, 
\ln |\tau-\tau^\prime|\, \nonumber \\
& &\times[ \dot{X}(\tau^\prime) \dot{Y}(\tau) +
\dot{Y}(\tau^\prime)  \dot{X}(\tau) ]\,\nonumber \\
& &\times\Big[ \int_0^1 d \xi 
\frac{ 2 \xi X(\tau) Y(\tau)}{[1+ \xi X^2(\tau)]^2}\,
\frac{1}{1+\xi X^2(\tau^\prime)} \nonumber \\
& &+
\int_0^1 d \xi \frac{1}{1+ \xi X^2(\tau)}\,
\frac{2 \xi X(\tau^\prime) Y(\tau^\prime)}{[1+ \xi X^2(\tau^\prime)} \Big].
\ea
\ba
& &S_{XY,9}=\frac{1}{\pi^2}\,\int d \tau d \tau^\prime \ln|\tau-\tau^\prime|\,
\dot{X}(\tau^\prime) Y (\tau)\,\nonumber \\
&\times& 
\int_0^1 d \xi 2 \xi  \frac{X(\tau) \dot{Y}(\tau)}{[1+ \xi X^2(\tau)]^2}\,
\frac{1}{[1+ \xi X^2(\tau^\prime)]}.
\ea
\ba
& & S_{XY,10}=\frac{1}{\pi^2}\,\int d \tau d \tau^\prime \ln|\tau-\tau^\prime|\,
\dot{Y}(\tau^\prime) X(\tau)\,\nonumber \\
& &\times\Big[ \int_0^1 d \xi  \frac{1}{[1+\xi X^2(\tau^\prime)]} Y(\tau)\,
\frac{d }{d \tau} \left( \frac{2 \xi X(\tau)}{[1+ \xi X^2(\tau)]^2} \right) 
\nonumber \\
& &+\int_0^1 d \xi \frac{d}{d \tau} \left ( \frac{1}{[1+ \xi X^2(\tau)]} \right )\,
\frac{2 \xi X(\tau^\prime) Y(\tau^\prime)}{[1+\xi X^2(\tau^\prime)]^2}\Big].
\ea

%%%%%%%%%%%%%%%%%%%%%%%%%%%%%%%%%%%%%%%%%%%%%%%%%%%%%%%%%%%%%%%%%%%%%%%%%%%%%%%%%%
\section{Derivation of RG equations}
We will closely  follow the original work by Anderson, Yuval, and Hamann,\cite{AGH}, so that  
only the major steps will be described below.
First  the trajectories with instanton-antiinstanton pairs in the interval
$[\tau_0, \tau_0+d \tau_0]$ are to be integrated out. Let $\delta Z$ be the part of partition function
containing only {\it one} close pair. The trajectories containing two or more pairs are accompanied with a factor 
of $(d \tau_0)^2$ and are neglected in our (essentially one-loop) approximation. 
\ba
\label{deltaZ}
\delta Z&=&\prod_i \Big[1+ z^2 \int_{t_i+2 \tau_0}^{t_{i+1}-\tau_0} d t^\prime \,
\int_{t^\prime -[\tau_0+ d \tau_0]}^{t^\prime-\tau_0} d t^{\prime \prime}  \nonumber \\
&\times& \exp[\delta V(t^\prime-t_1,\cdots,t^\prime- t^{\prime \prime})] \Big].
\ea
$t^\prime, t^{\prime \prime}$ denote the locations of  the close pair.
Writing out the exponent $V$ in detail 
\ba
\label{rg2}
\delta V&=&K_1 \sum_j(-)^{i+1+j}\,\ln \left | \frac{t^{\prime \prime}-t_j}{\tau_0} \right | \nonumber \\
&+&K_1 \sum_j (-)^{i+2+j}\,\ln \left | \frac{t^{\prime }-t_j}{\tau_0} \right |  \nonumber \\
&+&K_2 \sum_j\,\ln \left | \frac{t^{\prime \prime}-t_j}{\tau_0} \right | d(t^{\prime \prime}-t_j) \nonumber \\
&+&K_2 \sum_j\,\ln \left | \frac{t^{\prime }-t_j}{\tau_0} \right | d(t^{\prime }-t_j) \nonumber \\
&-&  K_1 \,
 \ln  \left | \frac{t^{\prime }-t^{\prime \prime }}{\tau_0} \right |+K_2 
 \ln  \left | \frac{t^{\prime }-t^{\prime \prime }}{\tau_0} \right |  d(t^{\prime }-t^{\prime \prime }). 
\ea
The last line of Eq.  (\ref{rg2}) can be neglected since $t^{\prime }-t^{\prime \prime } $ is the order of cutoff $\tau_0$.
The most important difference between $K_1$ part  and $K_2$ part of Eq.  (\ref{rg2}) is the 
absence of alternating sign factor in $K_2$ part.
With the mean value theorem applied for the integration over $t^{\prime \prime}$ of Eq.  (\ref{deltaZ})
one can effectively impose the relation $t^{\prime \prime}=t^\prime -\tau_0$  in other parts of integrand, 
and then one factor of $d \tau_0$ is multiplied.
Now the exponent $V$ becomes
\ba
\label{step1}
\delta V&=& K_1 \sum_j \, (-)^{i+j}\, \Big[ \ln \, \left | \frac{t^{\prime }-t_j}{\tau_0} \right |-
\ln \, \left | \frac{t^{\prime}-\tau_0-t_j}{\tau_0} \right |  \Big ] \nonumber \\
&+&K_2 \sum_j \,  \Big[ \ln \, \left | \frac{t^{\prime }-t_j}{\tau_0} \right | d(t^\prime-t_j) \nonumber \\
&+& \ln \, \left | \frac{t^{\prime}-\tau_0-t_j}{\tau_0}  \right | d(t^\prime-\tau_0-t_j)  \Big ].
\ea
We proceed by expanding $e^{\delta V} \sim 1+ \delta V$.
The Taylor expansion of the second term of $K_2$ part  of Eq.  (\ref{step1}) respect to $\tau_0$ 
gives
$$ \sum_j  \int d t^\prime  \partial_{t^\prime}   d(t^\prime- t_j)  = \sum_j \Big[ d(t_{i+1}-t_j)-  d(t_{i}-t_j) \Big],$$
which vanishes upon the summation over $i$ of the following approximation.
\be
\prod_i (1+ d \tau_0\,a_i) \sim 1+ \sum_i d \tau_0\,a_i+ o((d \tau_0)^2).
\ee
Collecting the remaining factors we obtain
\ba
\label{step2}
\delta Z&=&\prod_i \Big[1+ z^2 d \tau_0  \big \{ (t_{i+1}-t_i-3 \tau_0)  \nonumber \\
&+& \int^{t_{i+1}-\tau_0}_{t_i+2 \tau_0} d t^\prime 
[ K_1 \sum_j (-)^{i+j} \frac{\tau_0}{t^\prime-t_j} \nonumber \\
&+& 2 K_2 \sum_j 
\ln \, \left | \frac{t^{\prime }-t_j}{\tau_0} \right | d(t^\prime-t_j) ] \big \} \Big].
\ea
Re-exponentiating the expression in the  big bracket  of Eq.  (\ref{step2}) the correction to the partition function becomes

\ba
\label{step3}
& &\delta Z \sim \exp \Big[ z^2 d \tau_0 \sum_i  \big(   t_{i+1}-t_i -3 \tau_0 \big ) \nonumber \\
& & + z^2 d \tau_0\,K_1 \tau_0 \sum_{i\neq j} (-)^{i+j}
\ln \left| \frac{ t_{i+1}-t_j}{t_i-t_j} \right|   \nonumber \\
& &+2 z^2 d \tau_0\,  K_2 \sum_i \int_{t_i}^{t_{i+1}}\,d t^\prime \,  \sum_j 
\ln \, \left | \frac{t^{\prime }-t_j}{\tau_0} \right | d(t^\prime-t_j)   \Big ].
\ea
The first term in the exponent of Eq.  (\ref{step3}) contributes to
the normalization of free energy $e^{z^2 \beta  d \tau_0}$.
The $K_1, K_2$ part of Eq.  (\ref{step3})  behave very differently owing to the presence/absence of the oscillating factor.
\ba
\label{step4}
& &\delta Z \sim  \exp\Big[ z^2 d\tau_0 (K_1 \tau_0)  \sum_{i \neq j} (-)^{i+j} (-1) 2  \ln | t_i -t_j|  \Big ] \nonumber \\
& & \times \exp\Big[ z^2 d\tau_0 \{ \int_0^\beta d t^\prime \sum_j \,
 \big( 2 K_2  
\ln \, \left | \frac{t^{\prime }-t_j}{\tau_0} \right | d(t^\prime-t_j \big ) \}\Big ].
\ea
The $K_2$ part of Eq.  (\ref{step4}), in fact, does not depend on $t_j$ owing the integration over the whole range of
imaginary time. Thus, the $K_2$ part of Eq.  (\ref{step4}) is proportional to the number of instantons, and consequently
it  contributes only to the fugacity correction.
From the structure of Eq.  (\ref{step4}) 
the renormalization of $K_1$ is easily found
\be
\tilde{K}_1=K_1- 4 K_1  (z \tau_0)^2\,\frac{d \tau_0}{ \tau_0}.
\ee
The renormalization of fugacity  is given by
\ba
& &(\tau_0+ d \tau_0) \tilde{z}= (\tau_0 z) \frac{\tau_0+ d \tau_0}{\tau_0}\, \times \nonumber \\
& & \exp\Big[-\frac{K_1}{2}\,\ln \frac{\tau_0+d \tau_0}{\tau_0}  
  + z^2 \tau_0 d \tau_0 (2 K_2 g_2 )\Big],
\ea
where $g_2$ is
\be
\label{g2}
g_2=\int  \frac{d t^\prime}{\tau_0}  \ln \, \left | \frac{t^{\prime }}{\tau_0} \right | d(t^\prime).
\ee
The dependence of $d(t^\prime)$ is far stronger than $ \ln \, \left | \frac{t^{\prime }}{\tau_0} \right |$, thus we neglect the 
logarithmic factor up to logarithmic accuracy.
The explicit expression of $d(t^\prime)$ is 
\be
d(t^\prime ) = \frac{2 \lambda \gamma^2 \Delta}{\omega_0^2}\,\int_{-\infty}^\infty\,\frac{d \omega}{2\pi}\, 
\frac{\omega^2  e^{-i \omega t^\prime}}{\omega^2+ E^* |\omega| + \omega_0^2}.
\ee
Since $t^\prime \neq 0$ the above can be rewritten as
\be
d(t^\prime )=-\frac{2 \lambda \gamma^2 \Delta}{\omega_0^2}\, \int_{-\infty}^\infty\,\frac{d \omega}{2\pi}\,
\frac{(E^* |\omega| + \omega_0^2) e^{-i \omega t^\prime}}{\omega^2+ E^* |\omega| + \omega_0^2}.
\ee
The $E^* |\omega|$ is a small perturbation at both high and low energy. The correction coming from  $E^* |\omega|$
is the order of  $E^*/\omega_0 \ll 1$.
In leading approximation we can disregard  $E^*$.
Explicit evaluation gives
\be
d(t^\prime)=-\frac{\lambda \gamma^2 \Delta}{  \omega_0} e^{- \omega_0 t^\prime}.
\ee
The evaluation of $g_2$ gives
\be
\label{g2result}
g_2 \sim -\lambda \gamma^2 \Delta  \frac{e^{-\omega_0 \tau_0}}{\omega_0^2\tau_0}.
\ee
The scaling equation of fugacity  is  
\be
\label{scaling2}
\frac{d (z \tau_0)}{d \tau_0 /\tau_0}=(z \tau_0) \big[ (1-\frac{K_1}{2}) + (z \tau_0)^2 ( 2 K_2 g_2 ) \big].
\ee
It is convenient to introduce new variables which are more suitable 
in the context of original Kondo model.
\be
dl= d \tau_0 /\tau_0, \;\;  x= z \tau_0, \;\; K_1=2-y,\;\; y \ll 1.
\ee
In terms of new variables
the Eq.  (\ref{scaling2}) can be written
\be
\label{scaling3}
\frac{d x }{dl}=\frac{x}{2}\Big[y + x^2 \alpha  e^{-\omega_0 \tau_0} \Big].
\ee
In the Eq.  (\ref{scaling3}) the short time cutoff was identified with $1/\Delta$, 
while for the path integral approach the apparent cutoff is $1/U$.
This ambiguity has already been discussed in detail by Hamann in 
Sec. IV. B of Ref.[\onlinecite{hamann}], and it was argued that the ambiguity does not
give rise to any significant errors.

%%%%%%%%%%%%%%%%%%%%%%%%%%%%%%%%%%%%%%%%%%%%%%%%%%%%%%

%%%%%%%%%%%%%%%%%%%%%%%%%%%%%%%%%%%%%%%%%%%%%%%%%%%%%%%%%%%%
\end{document}